# Fine-Grain Authorization for Resource Management in the Grid Environment

K. Keahey
keahy@mcs.anl.gov
Argonne National Laboratory, Argonne, IL, USA

V. Welch
welch@mcs.anl.gov
U. of Chicago, Chicago, IL, USA

**Abstract**

*In this document we describe our work-in-progress for enabling fine-grain authorization of resource management. In particular we address the needs of Virtual Organizations (VOs) to enforce their own polices in addition to those of the resource owners[1].*

## 1   Introduction

In some Virtual Organizations (VOs) [1] the primary motivation for using the Grid is remote sharing of application services deployed on community resources, rather directly sharing those hardware resources themselves [2]. Since hardware resources are shared only through the agency of the VO, and the community is large and dynamically changing, the traditional mode of operation requiring each user to obtain an account from the owner of each resource participating in the VO is no longer satisfactory. Instead, we see an increasing trend to allow the use of both hardware and software resources based on VO credentials. In addition, sharing VO-owned application services requires VO-wide (as opposed to resource-specific) mechanisms for managing both these services and the VO's resource usage rights related to their execution. In this paper, we present an architecture relying on VO credential for service and resource management and allowing us to specify and enforce VO-wide service and resource usage policies.

We propose changes and extensions to the current Globus Toolkit's (GT2) resource management mechanism [3] to support this enforcement of rich VO policies. We describe how we can combine policies that are *resource-specific*, that is, determined by the resource owner, and *community-wide*, that is set by the VO. The goal of our work is to provide an architecture that allows a Virtual Organization to express policy on by whom and how its allocation is consumed, while at the same time ensuring that the resource provider's policies are still honored. Further, we consider two kinds of policy targets: application services, and traditional computing resources.

In the remainder of this paper we will propose a set of mechanisms enabling VOs to realize the scenarios described above for resource management. We discuss this architecture in the context of the current capabilities of the Globus Toolkit's (GT2)

[1] This work was supported by the Mathematical, Information, and Computational Sciences Division subprogram of the Office of Advanced Scientific Computing Research, U.S. Department of Energy, and by the Scientific Discovery Through Advanced Computing (SciDAC) Initiative

resource management mechanism [3] and propose extensions. We are currently implementing this architecture using GRAM and the Grid Security Infrastructure [4] mechanisms.

## 2   Use Scenarios and Requirements

VOs are often interested in setting policies on its member's use of community resources, not only in terms of *who* can use *what* resource, but *how* they are used. In this section we describe scenarios driving our work and illustrating the kinds of authorization required.

1. *Combining policies from different sources*. A supercomputing center decides to give allocations on some of its resources to several VOs specifying how much resource each of the VOs is allowed to use. The VOs can then decide: (1) which services can be run on these resources by which of their members and (2) how much of the community allocation can be used by individual members. This scenario requires combining policies coming from 2 different sources: the resource owner, and the VO. These policies can be expressed in different languages, and can contain dependencies necessary to resolve for full evaluation.

2. *Fine-grain control of how resources are used.* A VO has two groups of users: one group has the role of developing, installing and debugging the application services used by the VO to perform their scientific computation and the second group runs analysis using the application services. The first group may need a large degree of freedom in the types of applications they can run (e.g. compilers, debuggers, the applications themselves) in order to debug and deploy the VO application services, but should only be consuming small amounts of traditional computing resources (e.g. CPU, disk and bandwidth) in doing so. The second group may need the ability to consume large amounts of resources in order to perform science, but should only be doing so using application services approved by the VO. Furthermore, VO may wish to specify policies that certain users may use more or less resources than others and that certain applications may consume more or less resources than others.

3. *VO-wide management of jobs and resource allocations*. Currently, the only users who are allows to manage (e.g. suspend, kill) a job are: the user who instantiated it and any administrators on the resource on which it is running. However, for jobs using VO resource allocations it is often desirable for the VO to be able specify policy on who can manage a job. For example, users in a particular VO often have long-running computational jobs using VO resources and this same VO often has short-notice high-priority jobs that can't wait until other jobs are finished. Since it is often difficult to quickly find the users who submitted the original jobs, the VO wants to give a groups of it's members the ability to manage any jobs on VO resources so they can instantiate high-priority jobs on short notice.

Although we express our requirements in terms of authorization properties, and important aspect of our work deriving from these requirements involves creating enforcement mechanisms suitable for fine-grain authorization enforcement.

## 3 Problem Statement, Requirements, and General Approach

In order to support the use cases described in the previous section, we need to provide resource management mechanisms that allow the specification and consistent enforcement of authorization and usage policies set by a VO in addition to policies specified by the resource providers. In addition to allowing the VO to specify policies on standard computational resources, like processor time and storage, we want to allow the VO to specify policies on application services that it deploys as well as long-running computational jobs instantiated by community members.

In our work we will assume the following interaction model: An interaction is initiated by a user submitting a request, composed of the action of starting a job and the job's description, accompanied by the user's Grid credentials. This request is then evaluated against both resource and VO policies at different policy evaluation points (PEPs) located in the resource management facilities. If the request is authorized, it is carried out by local enforcement mechanisms operating based on local credentials. During the job execution, a VO user may submit management requests composed of a management action (e.g. request information, suspend or resume a job, cancel a job, etc.) In other words, following a pattern generally present in the Grid architecture, the enforcement module is an intermediary that translates grid-specific capabilities into local capabilities.

## 4 Overview of Current GRAM System

The current Globus Toolkit GRAM (Grid Resource Acquisition and Management) [3] system has two major software components: the Gatekeeper and the Job Manager. The Gatekeeper is responsible for creating a Grid service requested by the user. The Job Manager Instance (JMI) is a Grid service providing resource management and job control. This section will analyze the current system and explain its limitations.

### 4.1 Gatekeeper

The Gatekeeper is responsible for authenticating and authorizing a Grid user. The authorization is based on the user's grid credential and an access control list contained in a configuration file called grid-mapfile. This file is also used to map the user's Grid identity to a local account effectively translating the user's Grid credential into a local credential. Finally, the Gatekeeper starts up a Job Manage Instance (JMI), executing with the user's local credential. This mode of operation requires the user to have an account on the resource and implements enforcement only to the extent defined by privileges on this account.

### 4.2 Job Manager Instance (JMI)

The JMI parses the user's job startup request, and interfaces with the resource's job control system (e.g. LSF) to initiate the user's job. During the job's execution the JMI monitors its progress, and handles job management requests from the user. As the JMI is run under the user's local credential as defined by the user's account, the OS and local job control system are able to enforce local policy tied to that account on the JMI and user job.

The JMI does no authorization on job startup. However, once the job has been started, the JMI accepts, authenticates and authorizes management requests (e.g. suspend, stop, query, etc.) on the job. The authorization policy on these management requests is that the user making the request must be the same user who initiated the job. There is no provision for modifying this policy.

### 4.3 GRAM Shortcomings

The current GRAM architecture has a number of shortcomings when matched up with the requirements we laid out in Section 2:

1. Authorization of Grid service and user job startup is coarse-grained and not up to the expressiveness required.

2. Authorization on job management is coarse-grained and fixed to allow only the user who initiated a job to manage it.

3. Enforcement is implemented chiefly through the medium of privileges tied to a statically configured local account (JMI runs under local user credential) and therefore useless for enforcing fine-grained policy or policy coming from sources external to the resource (such as a VO).

4. Local enforcement depends on the rights attached to the user's account, not on the rights associated with a specific request and Grid credential accompanying that request.

5. A local account must exist for a user; this creates an undue burden on system administrators and users alike and prevents wide adoption of the network services model in large and dynamically changing communities.

These problems can, and have been, in some measure alleviated by clever setup. For example, the impact of (4) can be alleviated by mapping a grid identity to several different local accounts with different capabilities. (5) is often coped with by working with "shared accounts" (which however introduces many security, audit, accounting and other problems) or by providing a limited implementation of dynamic accounts [5].

## 5 Proposed Authorization and Enforcement Extensions to GRAM

In this section we describe our work in progress on implementing extensions to GRAM intended to overcome the shortcomings described above. Our works targets extensions to GRAM for policy evaluation including the design of a policy language for resource management, and strategies suitable for fine-grain policy enforcement.

### 5.1 Authorization System Extensions

Our requirements bring forth the need to replace the authorization methods currently used in GRAM by systems that are capable of evaluating complex fine-grain policies coming from multiple sources; in our case specifically the resource provider and the VO. We are currently working with two systems that meet these requirements: Akenti

[6] and the Community Authorization Service (CAS) [7]. Both of these systems allow for multiple policies sources, but have significant differences, both in terms of architecture (Akenti uses a pull model to query outside sources while CAS uses a push model where the user gets credentials from outside sources and pushes them to the resource) and programming APIs. We are in the process of experimenting with using either, or both of these systems (to combine different policy sources).

In order to retain flexibility in the choice of an authorization system, we defined a generic policy evaluation API that could be called by the PEP. This API will include passing, at a minimum, the following elements to the authorization system: the user's grid credentials, the name(s) of the policy target, and a description of the action the user is requesting.

### 5.2 Policy Language

GRAM allows user to start and manage jobs by submitting requests composed of an *action,* describing what is to be done with a job (start, cancel, provide status, change priority, etc.) and a job description. The job description is formulated in terms of attributes specified by the Resource Specification Language (RSL)[3]. RSL consists of attribute value pairs specifying job parameters such as executable description (name, location, etc.), and resource requirements (number of CPUs to be used, maximum allowable memory, etc.).

We are currently designing a policy language that allows for specification of the contents of the job description in terms of RSL and concepts related to job management such as actions, job ownership, and jobtags (see below). This allows a policy to limit not only the usage of traditional computational resources, but to dictate the executables they are allowed to invoke, allowing a VO to limit the way in which they can consume resources.

In order to specify VO-wide job management policies we introduce the notion of job tags. By requiring that a job have a certain jobtag we define a group of jobs that we can write policy about. This allows us to make policy about those jobs, for example to grant a set of users, who have an administrative role within a VO the right to manage those jobs. In order to implement it we extended RSL to accept a jobtag as a parameter; a VO user can then be required to submit a job with a specific jobtag (or any jobtag depending on the policy) and a user with administrative privileges can be given the right to use the jobtag to manage the jobs tagged by it. At present, jobtags are defined statically by a policy administrator, but we envision an approach in which the users will define them dynamically.

In the current implementation we experiment with the following assertions in our policy language:

- The job request can contain a particular attribute with the following values (e.g. enumerated list, range, regular expression or combination)
- The job request must/must not contain a particular attribute with one of the following values (e.g. it must specify the following queue, or a single processor, etc.)
- The job request must/must not contain a particular attribute (e.g. a jobtag must be specified)

- The job request must not contain any attributes not specified in this policy (in other words unless something is specified it is assumed to be forbidden)

So far we have found that these assertions cover the range of semantics we need to express.

### 5.3 Policy Enforcement

The current enforcement methods relied on by GRAM are unsuitable for enforcement of dynamically changing, fine-grained policies coming from sources external to the resource such as present in our requirements. In order to improve them we are working on thrusts in two areas: (1) implementing an enforcement gateway in GRAM itself and associated resource management tools, (2) implementing dynamic accounts, and sandboxing technologies. Combined together these approaches allow us to control external job initiation and management, securely support users who do not have an account on they system, and control locally operations of a job which we believe to be necessary and sufficient to securely implement our policies.

#### 5.3.1 Implementing enforcement in GRAM

Implementing enforcement in GRAM means creating a gateway controlling all external access to a resource; an action is authorized or not depending on decision yielded by a gateway. Policy can be enforced in GRAM at multiple PEPs corresponding to different decision domains; for example a PEP placed in the Gatekeeper can allow or disallow the creation of a Grid service. Since our work focuses on job and resource management we established a PEP in the JM where user requests are parsed and can therefore be evaluated. The PEP evaluates a request in the context of user credentials and policies from multiple sources and, if authorized, carries out the action. Since the PEP can deal with the full range of actions implemented by GRAM, it also allows users other than the initiator of a job to manage a job. We are working on a set of resource management tools to deal with situations where the user's local credential (carried by JMI) is not sufficient to carry out the request.

The weakness of this approach is that once the resource gateway decides to allow an action (for example a job execution), it has no control over subsequent actions of the job including actions specifically forbidden by the gateway. This places much responsibility in the hands of policy administrator, code developers and screeners, etc. who have to ensure that no undesirable actions will be taken by the code itself. In short, the gateway solution is similar to firewalls in that it places severe limitations on how initial connections to resources can be made, but unlike firewalls it depends on a wide range of variables that will be hard to control (such deep understanding of the implications of the actions of a complex code).

#### 5.3.2 Dynamic Accounts and Sandboxing

A sandbox is an environment that imposes restrictions on resource usage [8]. Sandboxing represents a strong enforcement solution, having the resource operating system act as the policy evaluation and enforcement modules and is largely complementary to the gateway approach. It is usually implemented by using platform-specific tools such as the Java Virtual Machine, or operating system specific capabilities. While they provide a solution with relatively high degree of security, they are hard to implement portably and introduce a degree of inconsistency across different platforms. At present our focus in this area is to tie our sandboxing needs to the dynamic accounts.

Dynamic Accounts are accounts that are created and configured on the fly by a resource management facility. This enables the resource management system to run jobs on a system for users that do not have an account at that system, and it also enables account configuration relevant to policies for a particular resource management request as opposed to a static user's configuration. Because of that, a dynamic account configuration can be also used as a sandbox on the user's rights. For example, by modifying user's group membership to control file system access, the account's quotas and other limits on resource usage, we can ensure that the user does not use more resources than authorized. On the other hand, accounts allow the user to modify only very few configuration parameters, and hence the enforcement implemented in an account is coarse-grained and may need to be supplemented by sandboxing technologies in order to implement fine-grained enforcement.

## 6  Summary

We described a work in progress aiming to provide mechanisms for VO-wide authorization and enforcement. The purpose of this work is to make VO-based trust model acceptable to resource owners and also to provide mechanisms enabling making and enforcing policies related to VO-wide operations such as resource management. Our system is designed to support fine-grain authorization on job startup and management, VO-wide job as well as resource allocation management. We are also working on strategies suitable for fine-grain policy enforcement